# The Economics of Augmented and Virtual Reality[*]

Joshua Gans
U Toronto and NBER
joshua.gans@utoronto.ca

Abhishek Nagaraj
UC Berkeley and NBER
nagaraj@berkeley.edu

Version 0.9: 26[th] May 2023

This paper explores the economics of Augmented Reality (AR) and Virtual Reality (VR) technologies within decision-making contexts. Two metrics are proposed: Context Entropy, the informational complexity of an environment, and Context Immersivity, the value from full immersion. The analysis suggests that AR technologies assist in understanding complex contexts, while VR technologies provide access to distant, risky, or expensive environments. The paper provides a framework for assessing the value of AR and VR applications in various business sectors by evaluating the pre-existing context entropy and context immersivity. The goal is to identify areas where immersive technologies can significantly impact and distinguish those that may be overhyped.

---

[*] Thanks to Ajay Agrawal, Steve Mann and Scott Stornetta for helpful discussions on an earlier draft of this paper. Nilo Mitra and Clara Vo provided excellent research assistance. Responsibility for all opinions remains our own.

# I. Introduction

Advances in mobile computing technology have heralded the rise of new interfaces that allow human beings to consume digital information in more immersive ways.[1] Two primary forms of such immersive technologies are Augmented Reality (AR) technologies and Virtual Reality (VR) technologies. AR technologies overlay virtual content onto the real world, enhancing the user's perception by blending physical and digital elements. VR technologies, on the other hand, create simulated environments through computer technology that immerses users in a three-dimensional, artificial world using a head-mounted display (HMD) or goggles. Both AR and VR technologies have seen massive interest and investment from the technology and venture capital community in the last decade. This interest is driven by seemingly endless applications of such technologies in a variety of different consumer and industrial contexts, including the potential development of the "metaverse" that creates shared experiences leveraging immersive technologies.

It is possible to define the scope of AR and VR technologies quite widely, and they need not be digital in nature. For instance, eye-glasses and a compass are examples of AR, while a ride like Disney's Pirates of the Caribbean could be considered a VR technology. Here we will, for the sake of containment, focus on digital forms of AR and VR technologies, usually in the form of some form of device worn across the eyes, coordinated with another (or the same) device worn on the ears and perhaps with a device involving haptic feedback (on the hands).

In the AR space, the first such device, the EyeTap, was built by Steve Mann in 1980. It consisted of a headset with a camera over one eye that displayed the wearer's surroundings and could be overlaid with text generated by a computer. At the time, it was a fully portable computer held in a backpack. However, later follow-on devices such as Google Glass were linked to a mobile phone for computational tasks. Mann also developed one of the first VR technologies; a large protective helmet (the MannGlas Welding Helmet) with a high dynamic range (HDR)[2] camera attached on the outside and a screen on the inside that allowed a welder to perform tasks in darkness and filtering the light distortion to see the area being welded very clearly.

We will also limit our examination on another dimension: to focus on AR and VR technology's usefulness in decision contexts. Many recent applications have focused on providing a more realistic and alternative setup for computer gaming, the ability to watch movies on a perceived large screen or simply a passive immersive experience.[3] Such applications have value, but our focus will be on applications that create value by shaping decision-making in business and consumer applications. This paper offers a framework *to identify the factors likely to shape the*

---

[1] The term "immersive technologies" is a catchall that has received widespread usage. For instance, "Immersive technologies create distinct experiences by merging the physical world with a digital or simulated reality." (https://www.vistaequitypartners.com/insights/an-introduction-to-immersive-technologies/).
[2] HDR was also invented by Mann (https://patents.google.com/patent/US5828793).
[3] For the avoidance of doubt, we also do not consider simple head-mounted devices such as GoPro and cameras on Google Glass or Snapchat's Spectacles as an immersive technology but simply sensors that happen to be worn in ways that may be similar to AR or VR technologies.

*value of opportunities created by AR and VR technologies*. We argue that the economics of AR and VR is best understood by studying how these technologies, as new tools that are interposed between humans and their environment, shape the value of two key activities humans use to navigate the world – accessing and being immersed in various contexts (environments) and understanding these contexts.

We define two dimensions that clarify these concepts. The first dimension is **Context Entropy** — a measure of the informational complexity of a given environment. The second is **Context Immersivity** — a measure of the value gained from directly accessing and being fully immersed in a particular environment. We argue that AR technologies are fundamentally about reducing entropy (i.e., making complex contexts easier to read and understand), while VR technologies are about making distant, risky, or expensive contexts easier to access and be fully immersed in them at higher resolution. Insofar as immersive technologies lower the costs of decoding and being immersed in a context, their value will be a function of pre-existing context entropy and context immersivity in different applications.

The paper explores the economics of Augmented and Virtual Reality using these two dimensions to gauge the appropriateness of immersive technologies for different solutions. Our simple two-dimensional framework can be used to evaluate various business application areas where immersive technologies are likely to be most impactful and offer the greatest value while identifying others that are perhaps more hype than substance.

## II. Background and Motivation

Augmented and Virtual Reality both have distinct approaches. AR technologies enhance the real world by adding information (in the form of digital data and images, usually viewed through stand-alone devices such as glasses, headsets, or mobile screens) to objects and places. On the other hand, VR technologies are about complete immersion in a virtual world, creating a purely digital experience viewed and sensed through a VR headset and wearable sensors. There are also variations of these two, called Extended Reality (ER)[4] or Mixed/Mediated Reality (MR). Recent developments include the widespread use of AR in mobile applications and the advancement of VR with high-quality headsets, resulting in improved graphics and motion tracking.

While these technologies have been around for some time, propelled by applications in gaming and supported by prototypes of headsets and wearables, the metaverse has given the impetus for investments and improvement on these early offerings in enterprise scenarios to derive greater business value from their use. But, beyond a pure technology play, enterprises and future entrepreneurs must understand how these technologies can be used within their businesses to drive long-term growth and stay relevant and competitive. This paper offers a framework to guide enterprises on the economics of immersive technologies by isolating the factors that identify impactful opportunities when pursuing a solution based on AR or VR.

---

[4] Mann, S., & Wyckoff, C. (1991). Extended reality. *Massachusetts Institute of Technology, Cambridge, Massachusetts*, 4-405; and Mann, Steve, Yu Yuan, Tom Furness, Joseph Paradiso, and Thomas Coughlin. "Beyond the Metaverse: XV (eXtended meta/uni/Verse)." *arXiv preprint arXiv:2212.07960* (2022).

To this end, we argue that one needs to explore two important dimensions, each tied to one of these two immersive technologies. AR takes a real-world context (environment) and overlays a variety of digital information on it to guide and help the user better navigate and interact with it. In other words, AR *clarifies* the context the user is in. And the value is greater when the context is complex, chaotic, confusing, or dynamic. If AR can reduce that complexity by providing *relevant* information in *real-time*, it will have added value in a decision context.

VR, on the other hand, offers full immersion in a certain type of environment, an environment that can be hard to reach, expensive, hazardous, or risky to create (or navigate) in real life. It reduces the stakes in decision-making by offering a cheaper way to evaluate options or experience complex, novel, or changing environments before committing to a solution in the real world.

We argue that VR technologies reduce the cost of accessing and being immersed in novel, hazardous, or difficult-to-reach contexts (environments), while AR technologies reduce the cost of understanding and navigating a given context. Working together, they clarify uncertainty and reduce mistakes in decision-making in business and social contexts.

The paper explores the economics of immersive technologies using these two dimensions to gauge the appropriateness and impact of using its two key technologies in various business scenarios. Our analysis is based on a formal framework we are developing in another companion paper. Section III defines and motivates the two key dimensions that underpin our framework. Section IV uses this framework to evaluate various business application areas where immersive technologies are likely to be most useful and offer the greatest value. Section V concludes.

## III. A Simple Economic Framework

To understand AR and VR through the lens of economics, we hark back to the earlier generation of technology — the penetration of the internet and the rise of what is now called Web 2.0 — that has transformed everything since 1995. The internet has commoditized communication, making it inexpensive and ubiquitous — something that is now taken entirely for granted. Similarly, the web has moved from a pure document-serving platform to one for interactive activities, enabling the "new economy" based on online services and removing the dependence on time and space for commercial and social interactions.

From an economic point of view, as argued by Goldfarb and Tucker,[5] the internet and the web have primarily reduced the cost of communication and search. And when such costs are lowered, it leads to greater usage of activities that need more communication and more search, and hence greater economic output. A corollary to this is that we also start to use these technologies in novel situations so that new, complementary activities arise not envisioned by the original designers. The essence of this insight is simple: when technologies reduce the cost of doing something, we see more of that activity and all the associated activities that rely on it to perform their function. This drives economic growth.

---

[5] Goldfarb, A, and Tucker, C. (2019). Digital Economics. *Journal of Economic Literature*, 57(1):3-43.

Similar arguments apply to machine intelligence. One of us has argued elsewhere[6] that machine intelligence is essentially a prediction technology whose primary effect is to reduce the cost of predicting outcomes. Thus, goods and services in businesses such as transportation, agriculture, and healthcare, among others, that rely on prediction as an input will see a reduction in costs with better outcomes. Such successes will lead to even more use of prediction technology — machine intelligence — in hitherto unused areas leading to overall economic growth. (Similarly, another paper[7] argued how blockchain technology reduces the cost of verification.)

Immersive technologies are a new information-gathering, processing, and rendering medium that interposes itself between users and their environment or context. This new medium can enhance or augment the context with additional data or even (re)create an environment in a virtual, digital form. AR and VR are the technologies used for these purposes, both working together to enhance and create contexts that enrich the human experience.

So, pursuing an approach like those referenced earlier, we ask: what are the factors whose costs is reduced? We have identified two dimensions, one for each of the two technologies, AR and VR, that can help guide which scenarios might deliver the best value. (It is important to note that we consider AR and VR to be technologies that support different aspects of immersion, and, indeed, XR, which mixes aspects of both, is probably how these two technologies will be used in practice.)[8]

*Reducing the cost of information processing*

AR technologies alter a user's perception of their immediate environment, offering various computer-generated sensory data to enhance their experience and interactions with that environment. This allows the user to navigate and interact with that environment more effectively and efficiently. So, being able to pick up *relevant* information easily makes it quicker to process and take decisions. For example, AR-based heads-up displays have been used in flying and, more recently, automobiles to reduce the mental effort of translating data from one screen via the brain to physical action, reducing distraction and errors. This aids pilots and drivers in making decisions like when to turn, change lanes or use alternative routes. In healthcare scenarios, AR can be used to visualize organs and map out veins and nerves, etc., useful for split-second decision making during complex on-site and remote surgery as well as when training medical personnel.

---

[6] Agrawal, A., J. Gans and A. Goldfarb (2019), *Prediction Machines: The Simple Economics of Artificial Intelligence*, Harvard Business Review Press: Boston.
[7] Catalini, C. and Gans, J. S. (2020). Some simple economics of the blockchain. *Communications of the ACM*, 63(7):80–90.
[8] The framework we present here bears some relationship to the more ambitious taxonomy provided by Steve Mann. Mann's 2-dimensional M-β plane, a two-dimensional taxonomy characterized by Mersivity (degree of immersiveness, also known as Mediality), and β ("Bits" as in the entropy of information theory in the Claude Shannon sense which can be analog or digital), which later gave way to the α-β (Reality-Virtuality) plane of XR (eXtended Reality); See Mann, S. & Wyckoff, C. "Extended Reality" MIT 4-405, 1991 also available at http://wearcam.org/xr.htm; Mann, S. "Mediated Reality...," *Presence Connect*, 2002 August 6, also available at http://wearcam.org/presence/; and Mann, S, Yuan, Y, Furness, T, Paradiso, J, and Coughlin, T, "Beyond the Metaverse: XV (eXtended meta/uni/Verse)," https://arxiv.org/pdf/2212.07960.pdf.

The common theme in these examples is that AR delivers information more efficiently than done by conventional means, such as manuals or separate displays. AR technologies can be seen as a way to reduce cognitive load, allowing the user of these technologies to expend less effort to process information. These technologies can also simplify or add to the information in a given context in smart ways to make it easier to navigate environments. Thus, the productivity of any decisions that require environmental examination and interaction is increased. The success of such examples, in turn, will encourage more use of AR technologies in novel and unanticipated scenarios.[9]

We have formalized this lowering of the cost of obtaining information (or lowering the effort to process information) as **Context Entropy** — a measure of the informational complexity of the environment. Context refers to the environment under observation (including the subject interacting with it), while entropy is a term borrowed from information science to quantify the degree of uncertainty, randomness, change, or lack of predictability in that context. For example, a meeting between participants who know each other has low context entropy, while one between unknown persons has high context entropy. Consider an AR technology that can identify the individual a person is interacting with. In the absence of an AR technology, the user may not know the person at all, may not know that they know the person, or may identify the person but not with high confidence. An AR-based "virtual badge" identifies the person and provides their name and possibly other relevant information. This saves the user cognitive effort and helps then with the decision or whether and how to approach an individual for the purposes of networking. For instance, the user must no longer struggle with their memory or make sure they can see the face of the person when making this decision (e.g., they may be wearing a mask during a pandemic).

Thus, the AR technology provides information with a reduced impact on the user's cognitive attention. This allows them to expend that attention elsewhere, e.g., in surfacing important details about the identified person, such as their organization and role, although this too can potentially be AR-assisted.

It is useful, however, to point out what types of information that may be delivered through eye-wear or a heads up display would not provide value in reducing entropy. For instance, many AR head-wear (such as Google Glass) displayed notifications, the time and other information directly from a phone but in the eyeline of the wearer. It was argued that the user could consume the information without having to break eye contact in social situations – as might occur with looking at a phone or a watch. However, none of this information is helping the user in dealing with context entropy. Indeed, almost by definition, it is a distraction and will increase the costs to the user of processing contextual information.

---

[9] A critical aspect of using AR for increasing economic value by aiding decision-making is that the enhancements they generate reflect environmental truths. We use this to distinguish the productive use of AR from those used in certain consumer applications such as gaming or photography which alter or distort appearances or features – that would be *altered*, rather than augmented, reality. This is a subtle concept. For instance, sunglasses are an AR technology because they allow people to see things more clearly.

*Reducing the cost of immersion*

The second dimension is **Context Immersivity** – a measure of the value of accessing and being fully immersed in a particular environment to interact effectively with it. Immersivity has several additional properties beyond the obvious one of accessing complex, uncertain, hazardous, or risky environments.[10]

One is the psychological and physical aspects of being in an environment, with the ability to sense it physically and emotionally. An example might be the experience of being present at a rock concert versus watching a recording of it, no matter how well-crafted afterwards. Or the difference between seeing or reading a description or image of some complex object on a page — its 2D rendering, as it were — and touching and manipulating a 3D model with one's hands. Our mind processes these differently, and the latter is the one that has the greater impact.

This intangible aspect of immersivity — the psychological aspects of "being there" — captures why being immersed in an environment is important for decision-making, namely the use of subtle indicators of communication such as body language or slight eye movements that enable us to decide on actions in various social and interpersonal contexts. It will be a long way before virtual meetings can successfully capture such non-verbal aspects of human interactions.

To this end, it is useful to point out that many aspects of VR design – such as providing a virtual meeting room with interesting views or with participants with legs – may not provide an expansion of information relevant for decision-making. Indeed, these attempts at skeuomorphic design may trade-off against other dimensions that expand the set of information available to users.[11]

That said, it is not always possible to know what any person might need in a particular context to be able to interact successfully within it. VR offers an inexpensive way to recreate an environment acknowledging the fact that designers of systems and solutions do not exactly know how different people will react to the same context or know what someone needs from a given context. By replicating an environment using VR (and augmenting it with more information using AR), different people can explore and experience whatever is most useful and relevant to them.

---

[10] In a sense this is a very limited focus on immersivity. Specifically, it is about the information that can be provided about context that is potentially useful to decision-makers. It is a subset of a more general concept of "mersivity" (https://mersivity.com/) that includes concepts that break the human external/internal barrier. Here we focus exclusively on sensory information (that is, external).

[11] There are actually good reasons to suppose that perfectly replicating the environment (with all sensory impacts included) is not, in fact, the correct design goal for these technologies; see Hollan, J. and Stornetta, S., 1992, June. Beyond being there. In *Proceedings of the SIGCHI conference on Human factors in computing systems* (pp. 119-125). It is for this reason that we speculate that the optimal design on immersive technologies in various contexts will be a combination of AR and VR elements.

Finally, being in a given environment is not always practicable, owing to many factors — distance, physical capability, cost, risk, and so forth. VR can reduce the cost of accessing such environments allowing the user to experience and be fully immersed in them while making decisions. There are many instances when the cost of creating a model or prototype is very high (e.g., aerospace, automotive, construction), but the ability to experience and evaluate it is crucial for the success of the eventual product or solution. VR can be used to aid planning and design, leading to better decision-making before locking a customer or stakeholders into a solution that cannot easily be altered without exorbitant cost or delay (a car or airplane, say) or which is expected to remain unchanged for a lengthy period (such as a building).

VR, therefore, provides the greatest value when it reduces the cost of accessing contexts (environments) that are complex, uncertain, fantastical, hazardous, or risky. Thus, a virtual reality model of the Martian surface can allow planning a rover's route past obstacles, which has greater economic value and consequences than, say, exploring the layout of a rental apartment.

*A two-dimensional matrix of possibilities*

The remainder of this paper explores the economics of AR and VR technologies using these two dimensions to gauge the impact and appropriateness of using these two key technologies in various usage scenarios.

This will be expressed schematically as a two-dimensional matrix, with the rows identifying low and high context entropy while the columns represent low and high context immersivity. (Of course, there is a continuum between low and high, but the general idea is sufficient for illustrating our thesis.)

For example, an environment with high context entropy but low immersivity is a virtual office or conference scenario on a global scale that provides the exchange of trusted identities and background information using AR, allowing for participants to easily choose the right people to interact with. On the other hand, a VR model of the Martian surface — an application with high context entropy and high context immersivity — that allows planning a rover's route past obstacles will have greater economic value and consequences than exploring, say, the layout of a rental apartment which offers low context immersivity and entropy.

We argue that exploiting AR is about clarifying context entropy (i.e., making complex contexts easier to read and understand) while VR is about making it easier to access and be immersed in distant, hazardous, or expensive contexts (e.g., travel on Mars) with the latter including contexts that do not exist in the real world.

The following section analyzes different usage scenarios to identify how they are positioned along these two dimensions. We shall see that the value of immersive technologies will be highest in areas where it provides access to high-value and hard-to-navigate/novel contexts (e.g., designing the architecture for a new building or flight simulators) and low when not (e.g., we argue that early offerings such as virtual office meetings provide low value for AR and VR).

## IV. The Economic Value of Different Applications

In the previous section, we offered a framework to evaluate the value provided by AR and VR technologies using the two dimensions — context entropy and context immersivity — identified in Section III. In both cases, the value of using these technologies (or a mixture) lies in lowering the cost of decision-making in real-world applications.

A new generation of entrepreneurs has moved beyond gaming and entertainment and seeks applications in industries such as education, construction, automotive, healthcare, manufacturing, etc., to find compelling usage scenarios for important enterprise activities that can leverage AR/VR to improve efficiencies and reduce costs. In the following subsections, we evaluate, using our framework and these two dimensions, several usage scenarios chosen from these industry sectors.

*Low Context Immersivity – High Context Entropy*

We begin with decisions in contexts where immersivity is not highly valued but there is high context entropy. This is a use case favoring the use of AR but not VR technologies. An example of this would be a conference where participants must network but do not know each other making the context harder to parse. AR can provide value in such a context by identifying individuals by name, affiliation, and other (presumably privacy-protected) data, which can reduce friction in communication and encourage more spontaneous collaboration.

Another scenario is when people operate with complex equipment or navigate a complex or unfamiliar environment. This might also arise in dangerous situations like firefighting or military contexts. Appropriately placed AR technologies can provide key information allowing the user to process the information from the context more effectively and make split-second (and life-or-death) decisions.

Another scenario in this quadrant is for training personnel to use new equipment or navigate a complex environment, for example, a factory floor with moving objects and complex safety requirements, as well as the need to operate machinery and assemble products. Appropriately placed AR-based information viewed on heads-up displays or handhelds can guide users in such environments. These may also provide clearer training experiences, replacing, where possible, training manuals, classes, and on-site instructors.

*High Context Immersivity – Low Context Entropy*

These scenarios are ones where the cost of creating an environment is high and design mistakes can be expensive or even deadly; so, models and prototypes are needed to recreate and experience the environment before costly decisions are made.

This is particularly true in the fields of architecture and building construction, especially for commercial real estate, where the process of designing and constructing new structures is complex and time-consuming; so, the use of VR to plan and build out fully equipped "digital twins" of buildings (including furnished floors, placement of HEVC and security systems, down to the smallest details such as power outlets and light switches) can avoid painful surprises for structures that are expected to last decades. A similar scenario is greenlighting the design of a new car or aircraft, where a VR-created prototype allows potential consumers to experience the equipment as if they were in a showroom.

Such scenarios have low context entropy because the use of tools, equipment, materials, and processes used in such activities are usually well-known or change relatively infrequently. The value comes from high context immersivity – finding out what works because it is too costly to get it wrong in the real world and where the value of being psychologically immersed in a context is quite high.

Such use cases for VR technologies are particularly strong when users are being trained for real-world situations that involve high stakes. This was realized decades ago with significant VR investments in flight simulators to train airline pilots. However, it could be imagined that purely digital VR environments might be of use for the training of military, emergency and rescue workers or medical professionals (including surgeons) in making high-stakes decisions while being immersed in a given context.

*High Context Immersivity – High Context Entropy*

Certain decision scenarios have both high context entropy and require high context immersivity. These might include cases where there is a need to reach – or simulate using VR – a distant or hazardous location or both (e.g., the surface of Mars). At the same time, the environment under observation might offer unanticipated or unexpected challenges related to terrain or weather, or other random and dynamic factors, making these high entropy contexts.

AR can be particularly helpful when navigating other high-context entropy environments where there are higher stakes, such as life or death for errors, as might happen in medical scenarios. AR/VR can be used to aid with training, educating, and immersing both patients and healthcare professionals in medical information by superimposing digital scans (e.g., an MRI) on a patient's body to identify affected organs easily.

AR/VR can be used to aid doctors and surgeons when diagnosing and treating patients remotely. Treatment might involve telesurgery with remote collaboration via the HoloLens headset, guiding on-site surgeons. And training doctors in complex surgical procedures – possibly using robotic equipment manipulated remotely – would also sit in this quadrant.

On an interplanetary scale, a virtual reality model of the Martian surface can allow planning a rover's route past obstacles, which has greater economic value and consequences than exploring, say, the layout of a rental apartment.

*Low Context Immersivity – Low Context Entropy*

These are scenarios that are easy to navigate – low context entropy – and where it is not necessary to fully (re)create an environment – that is, low context immersivity – because the value gained from a fully immersive experience is low. In these contexts, the value for both AR and VR technologies may be small.

A virtual (or remote) meeting among known participants in an enterprise is a good example of a use case that scores low on both measures. Given the variety of remote collaboration tools with audio/video support and private/public chat, accessibility is reduced to good internet connectivity. It can also be argued that known participants, who have collaborated face-to-face previously, can navigate the environment best, being attuned to each other's verbal and non-verbal cues.

Importantly, it is unlikely that adding avatars to represent participants or finding ways to add physical cues such as smell, or touch can increase the value of this by several factors from what is available from current collaboration tools to make it a worthwhile exemplar for VR/AR.

The same might be said for virtual offices and classrooms. For the latter, there might be some value to getting the feeling of "being there", but the added value to learning is low.

*Summary*

Our analysis of various usage scenarios taken from a variety of enterprise sectors can be summarized in the following table. An example of a decision is presented in parentheses, along with the decision-making context.

|  |  | Context Immersivity | |
|---|---|---|---|
|  |  | Low | High |
| Context Entropy | Low | • Virtual Classrooms *(deciding the pace of instruction)*<br>• Virtual Offices and Workspaces (*choosing a colleague to get feedback from*)<br>• Virtual meetings with colleagues *(deciding a course of action after a meeting)* | • Building, construction, and architecture *(approving a design)*<br>• Training for emergency situations (e.g., firefighting, defense) *(issuing a certification)*<br>• Virtual tours, concerts, showrooms etc. *(assessing quality of the product/service)* |
|  | High | • Virtual Conferences *(choosing a new person to network with)*<br>• Training on new equipment *(deciding when to repeat a lesson vs. moving on)*<br>• Heads-up displays for navigation, training *(choosing among alternate routes/modules)* | • Prototyping and testing new equipment (e.g., aircraft, cars) *(approving a prototype for production)*<br>• Exploration of hazardous, inaccessible, or novel environments (e.g., mining sites) *(design of safety measures)*<br>• Remote (robot-aided) surgery *(choosing among alternate surgical options)* |

Our placement of usage scenarios in these quadrants is an initial attempt at using this framework to see where the highest business impact might be achieved. Most of these scenarios are currently constrained by the lack of suitable hardware (such as wearables and displays) for a truly immersive experience at a suitable price and comfort level. Their positioning in the quadrants may change depending on how the cost, versatility, and availability of these technologies evolve to offer better immersive experiences.

## V. Conclusion

Immersive technologies such as AR and VR offer a new information-gathering, processing, and rendering medium that interposes itself between users and their environment or context. We have provided a way to model the economics of such technologies, showing where value is added by using AR and VR in business contexts for improved decision-making.

We argue that the rise of immersive technologies involves a change in the cost of two key activities central for humans when navigating their environment (contexts): accessing contexts and understanding contexts. VR technologies reduce the cost of accessing and being immersed in new contexts, while AR technologies reduce the cost of understanding and navigating a given context. Using this framework, we have analyzed and identified a variety of business applications where metaverse technologies are likely to be fundamental and other areas where their impact is less substantial.

Our analysis shows that the value of AR and VR will be highest in areas where it provides access to complex, high-value, hard-to-access, and novel contexts (e.g., designing the architecture for a new building or planning and navigating a rover's path on Mars) and low when not (e.g., we suggest that early offerings such as virtual meeting rooms are a poor use case for showing the value of immersive technologies).